\documentclass[english, showpacs]{revtex4-1}
\usepackage{float}
\usepackage{amsmath}
\usepackage{amssymb}
\usepackage{graphicx}
\usepackage{esint}

\makeatletter

\newcommand{\lyxdot}{.}

\makeatother

\usepackage{babel}
\begin{document}

\title{The Master Equation for Two-Level Accelerated Systems at Finite Temperature}

\author{J.L. Tomazelli}

\email{jeferson.tomazelli@ufsc.br}

\affiliation{Departamento de Fisica, Universidade Federal de Santa Catarina, Caixa
Postal 476, CEP 88010-970, Florianopolis, S.C., Brasil}

\author{R.O. Cunha}

\email{renan.oliveira@pgfsc.ufsc.br }

\affiliation{Departamento de Fisica, Universidade Federal de Santa Catarina, Caixa
Postal 476, CEP 88010-970, Florianopolis, S.C., Brasil}

\thanks{corresponding author.}

\begin{abstract}
In this work we study the behaviour of two weakly coupled quantum
systems, described by a separable density operator; one of them is
a single oscillator, representing a microscopic system, while the
other is a set of oscillators which perform the role of a \emph{reservoir}
in thermal equilibrium. From the Liouville-Von Neumann equation for
the reduced density operator, we devise the master equation that governs
the evolution of the microscopic system, incorporating the effects
of temperature via Thermofield Dynamics formalism by suitably redefining
the vacuum of the macroscopic system. As applications, we initially
investigate the behaviour of a Fermi oscillator in the presence of
a heat bath consisting of a set of Fermi oscillators and that of an
atomic two-level system interacting with a scalar radiation field,
considered as a \emph{reservoir}, by constructing the corresponding
master equation which governs the time evolution of both sub-systems
at finite temperature. Finally, we calculate the energy variation
rates for the atom and the field, as well as the atomic population
levels, both in the inertial case and at constant proper acceleration,
considering the two-level system as a prototype of an Unruh detector,
for admissible couplings of the radiation field.
\end{abstract}
\maketitle

\section{introduction}

The interest in the study of quantum many-body systems from the modern
standpoint of quantum field theories \cite{Matsubara1955,Schwinger1961,Keldysh1964,Takahashi1975}
has grown since the 1960's , ranging from low to extremely high energy
and temperature regimes \cite{Umezawa1982,Kapusta2006,Das1997,Khanna2009,Tomazelli2003}.
In 1975, Takahashi and Umezawa proposed an operator-based formalism
called Thermofield Dynamics (TFD) \cite{Umezawa1995} in order to
describe quantum systems at finite temperature in thermal equilibrium.

In the same year the discovery of the Hawking effect \cite{Hawking1975,Hawking1974},
from the black hole thermal radiation, was one of the most innovative
results in quantum field theory in curved space-time. Later on, Davis
\cite{Davies1975} suggested that a similar effect would occur even
in a flat space-time. In 1976, Unruh \cite{Unruh1976} showed that
the essential characteristics of the Hawking effect were contained
in a simpler situation: an accelerated detector, would be excited
by particles in vacuum, property known as the Unruh effect. This discovery
of the Hawking-Unruh effect has inspired many researchers, generating
since them a series of important results, \cite{Davies1978,Sciama1981,Birrell1982}.
There are two traditional approaches to the Unruh effect, one is through
the quantization of non-massive field in curvilinear coordinates,
the other, due to DeWitt \cite{DeWitt1975} and which we shall use
in this work, is considering the acceleration as a hyperbolic parametrization
in Minkowski space time. 

The present work aims at investigating the manifestation of the Unruh
effect, approaching the issue from yet another perspective \cite{Audretsch1994,Audretsch1995}.
We consider a two-level system in interaction with a non-massive scalar
field, for which we deduce a master equation from Von Neumann equation
. Subsequently, we present the thermal states of Thermofield Dynamics
(TFD) in the number representation and use it to thermalize the theory
via vacuum expected values​​, singling out in a clearer way the Unruh
temperature.

The original proposal of Unruh and DeWitt, considers a detector as
a point object that is linearly coupled with a zero mass scalar field
through an interaction of monopole type

\begin{equation}
H_{I}\left(\tau\right)=\mu M\left(\tau\right)\phi\left(t\left(\tau\right),\mathbf{x}\left(\tau\right)\right)\mbox{,}\label{eq:Hamiltoniano.intera=0000E7ao.DeWitt}
\end{equation}
where $\mu$ is a coupling constant and $M\left(\tau\right)$ is the
monopole moment of the detector. Specifically, $H_{D}$ is the Hamiltonian
of the detector which has a discrete spectrum $\left\{ E_{i}\right\} .$
We assume that the detector moves along the world line described by
the coordinate $x\left(\tau\right)$, where $\tau$ is the proper
time of the detector. Assume that at an initial instant $\tau_{0}$
the detector is in the state $\left|E_{0}\right\rangle $ and the
field in the vacuum $\left|0\right\rangle $. For an arbitrary trajectory,
it is reasonable to assume that the detector will not remain in the
initial state, but undergoes a transition to an excited state. Therefore,
for a later time $\tau>\tau_{0}$, the detector is in the state $\left|E_{i}\right\rangle $
and field in the state $\left|\Psi\right\rangle $. For $\mu$ sufficiently
small, we can calculate via first order perturbation theory the transition
probability

\begin{equation}
\mathcal{P}_{t}=\mu^{2}\underset{i}{\sum}\left|\left\langle E_{i}\left|M\left(0\right)\right|E_{0}\right\rangle \right|^{2}\intop_{-\tau+\tau_{0}}^{\tau-\tau_{0}}du\,e^{-iu\Delta E}\left\langle 0\left|\phi\left(u\right)\phi\left(0\right)\right|0\right\rangle \mbox{.}\label{eq:probabilidade.transi=0000E7ao.total.final}
\end{equation}

The Wightman function in the integrand is well known and can be explicitly
calculated, giving

\begin{equation}
\left\langle 0\left|\phi\left(t,\mathbf{x}\right)\phi\left(t',\mathbf{x}'\right)\right|0\right\rangle =\frac{1}{4\pi^{2}\left[\left|\mathbf{x}\left(\tau'\right)-\mathbf{x}\left(\tau''\right)\right|^{2}-\left(t\left(\tau'\right)-t\left(\tau''\right)-i\epsilon\right)^{2}\right]}.\label{eq:funcao.Wightman}
\end{equation}

Firstly, we consider the case in which the detector follows a inertial
world line, i.e., the parametric equations

\begin{equation}
\begin{array}{c}
t\left(\tau\right)=\gamma\tau\mbox{,}\\
\\
\mathbf{x}\left(\tau\right)=\mathbf{x}_{0}+v\gamma\tau\mbox{,}
\end{array}
\end{equation}
from which we obtain for the Wightman function \eqref{eq:funcao.Wightman},

\begin{equation}
\left\langle 0\left|\phi\left(t,\mathbf{x}\right)\phi\left(t',\mathbf{x}'\right)\right|0\right\rangle =-\frac{1}{4\pi^{2}\left(\tau'-\tau''-i\epsilon'\right)^{2}+\mathcal{O\left(\mbox{\ensuremath{\epsilon^{2}}}\right)}}\mbox{,}
\end{equation}
and so the probability 

\begin{equation}
\mathcal{P}_{t}=-\frac{\mu^{2}}{4\pi^{2}}\underset{i}{\sum}\left|\left\langle E_{i}\left|M\left(0\right)\right|E_{0}\right\rangle \right|^{2}\intop_{-\infty}^{\infty}du\,\frac{e^{-iu\Delta E}}{\left(u-i\epsilon'\right)^{2}}\mbox{.}\label{eq:prob.transi=0000E7ao.inercial}
\end{equation}

The integral in \eqref{eq:prob.transi=0000E7ao.inercial} turns out
to be $\mathcal{P}_{t}=0$. Therefore, the transition probability
to an excited state is zero, since, as expected, a detector at rest
must not spontaneously excite. Now, consider the case where the detector
is uniformly accelerated; this is equivalent to a hyperbolic world
line, parametrized according to

\begin{equation}
\begin{array}{c}
x\left(\tau\right)=\alpha^{-1}cosh\left(\alpha\tau\right)\mbox{,}\\
\\
t\left(\tau\right)=\alpha^{-1}sinh\left(\alpha\tau\right).
\end{array}\label{eq:param.acelerada}
\end{equation}

Similarly, for the Wightman function,

\begin{equation}
\left\langle 0\left|\phi\left(\tau\right)\phi\left(\tau'\right)\right|0\right\rangle =-\frac{1}{4\pi^{2}}\underset{n=-\infty}{\overset{\infty}{\sum}}\frac{1}{\left[\left(\tau-\tau'\right)-i2\epsilon+i\frac{2\pi}{\alpha}n\right]^{2}}\mbox{,}\label{eq:Wightman.somatorio}
\end{equation}
so that the resulting probability is 

\begin{equation}
\mathcal{P}_{t}=-\frac{\mu^{2}}{4\pi^{2}}\underset{i}{\sum}\left|\left\langle E_{i}\left|M\left(0\right)\right|E_{0}\right\rangle \right|^{2}\underset{n=-\infty}{\overset{\infty}{\sum}}\intop_{-\infty}^{\infty}du\,\frac{e^{-iu\Delta E}}{\left[u-i2\epsilon+i\frac{2\pi}{\alpha}n\right]^{2}}.\label{eq:prob.transi=0000E7ao.acelerado}
\end{equation}

After integration, it turns out to be 

\begin{equation}
\mathcal{P}_{t}=\frac{\mu^{2}}{2\pi}\underset{i}{\sum}\left|\left\langle E_{i}\left|M\left(0\right)\right|E_{0}\right\rangle \right|^{2}\frac{\Delta E}{e^{\frac{2\pi\Delta E}{\alpha}}-1}\mbox{,}\label{eq:prob.transi=0000E7ao.rindler}
\end{equation}
from which one recognizes the factor $\left[e^{\frac{2\pi\Delta E}{\alpha}}-1\right]^{-1}$
as being the Planck distribution associated with the Unruh temperature

\begin{equation}
T=\frac{\alpha}{2\pi k_{B}}\mbox{,}\label{eq:temperatura.acelera=0000E7ao}
\end{equation}
where $k_{B}$ is the Boltzmann constant.

\section{Thermofield Dynamics\label{sub:teoria.vac.termico}}

In this section we briefly present the formalism introduced by Takahashi
and Umezawa \cite{Takahashi1975} for a real time thermal field theory,
where statistical averages of physical observables correspond to the
expected values​​ in a thermal Bogoliubov vacuum.

In Schwinger's Measurement Algebra (SMA) an operator is defined as

\begin{equation}
X=\underset{n,m}{\sum}M\left(n\right)XM\left(m\right)\mbox{,}
\end{equation}
where $M\left(n\right)=\left|n\right\rangle \left\langle n\right|$
is known as the measurement symbol or projector into the $n$ basis;
defining the composition rule

\begin{equation}
M\left(n\right)M\left(m\right)=\delta\left(n,m\right)M\left(m\right)\mbox{,}
\end{equation}
in such a context, the expectation value of a given observable $A$
in the single $m$ basis is the scalar

\begin{equation}
\left\langle A\right\rangle _{m}=Tr\left[AM\left(m\right)\right]\mbox{.}
\end{equation}

From that one, we can define the statistical average

\begin{equation}
\left\langle A\right\rangle =Tr\left[\rho A\right]=\underset{m}{\sum}\pi\left(m\right)\left\langle A\right\rangle _{m}\mbox{,}
\end{equation}
where $\pi\left(m\right)$ are the statistical weights and we can
recognized the density operator in the SMA as

\begin{equation}
\rho\equiv\underset{m}{\sum}\pi\left(m\right)M\left(m\right)\mbox{.}
\end{equation}

It is known that the inclusion of temperature may cause the duplication
of the system degrees of freedom \cite{Umezawa1995}. We can achieve
this by the introduction of the relation

\begin{equation}
\delta\left(n,m\right)=\delta\left(\widetilde{n},\widetilde{m}\right)\mbox{;}
\end{equation}
once this is done, we can rewrite the density operator as 
\begin{equation}
\begin{array}{c}
\rho=\underset{n,m}{\sum}\sqrt{\pi\left(n\right)\pi\left(m\right)}\delta\left(n,m\right)M\left(n,m\right)=\underset{n,m}{\sum}\sqrt{\pi\left(n\right)\pi\left(m\right)}M\left(n,\widetilde{n}\right)M\left(\widetilde{m},m\right)\\
\\
=\left[\underset{n}{\sum}\sqrt{\pi\left(n\right)}M\left(n,\widetilde{n}\right)\right]\left[\underset{m}{\sum}\sqrt{\pi\left(m\right)}M\left(\widetilde{m},m\right)\right]\equiv\left|0_{\left(\beta\right)}\right\rangle \left\langle 0_{\left(\beta\right)}\right|\mbox{,}
\end{array}
\end{equation}
where we have made use of the composition law of composite measurement
symbols. In thermal equilibrium, the statistical weights assume the
form $\pi\left(m\right)=Z^{-1}e^{-\beta E_{m}}$. In the above expression,
we can interpret the density operator as a projector into a thermal
vacuum state $\left|0_{\left(\beta\right)}\right\rangle $.

For a quantum system in thermal equilibrium at temperature $T$, in
the canonical ensemble, the statistical average of an observable A
is then given by 
\begin{equation}
\left\langle A\right\rangle \equiv Z^{-1}Tr\left[\rho A\right]=\left\langle 0_{\left(\beta\right)}\left|A\right|0_{\left(\beta\right)}\right\rangle \mbox{,}\label{eq:media.estatistica}
\end{equation}
where
\begin{equation}
\left|0_{\left(\beta\right)}\right\rangle =Z^{-\frac{1}{2}}\underset{n}{\sum}e^{-\frac{1}{2}\beta E_{n}}\left|n,\widetilde{n}\right\rangle \mbox{,}\label{eq:vac.termico.final}
\end{equation}
\[
\left|n,\widetilde{n}\right\rangle \equiv\left|n\right\rangle \otimes\left|\widetilde{n}\right\rangle \mbox{.}
\]

In order to construct this representation it's appropriated to introduce
a fictitious system identical to the original system, so that $\left|\widetilde{n}\right\rangle $
belongs to this new system. For the free scalar field, we have

\begin{equation}
\left|0_{\left(\beta\right)}\right\rangle _{\mathbf{k}}=\left[1-e^{-\beta\omega_{\mathbf{k}}}\right]^{\frac{1}{2}}\underset{\mathbf{k}=0}{\overset{\infty}{\sum}}e^{-\frac{1}{2}\beta\omega_{\mathbf{k}}n_{\mathbf{k}}}\left|n_{\mathbf{k}},\widetilde{n}_{\mathbf{k}}\right\rangle \mbox{,}
\end{equation}
from which follows the Bose-Einstein distribution

\begin{equation}
\left\langle 0_{\left(\beta\right)}\left|N_{\mathbf{k}}\right|0_{\left(\beta\right)}\right\rangle =\left[e^{\beta\omega_{\mathbf{k}}}-1\right]^{-1}\mbox{.}\label{eq:est.bos.einstein}
\end{equation}

\section{The Master Equation\label{sec:The-Master-Equation}}

In this section we outline the steps leading to the finite temperature
master equation following closely the procedure presented in \cite{Cohen-Tannoudji1992}
(see also \cite{Ford1988,Halliwell1996,Ford2001,Hu1992}). Consider
a microscopic system $A$ surrounded by a dissipative environment
$R$. The typical Hamiltonian that describes the total system, in
the Schrödinger representation, is 
\begin{equation}
H^{S}=H_{A}^{S}+H_{R}^{S}+V^{S}\mbox{,}
\end{equation}
where $H_{A}^{S}$ is the free Hamiltonian for the system $A$, $H_{R}^{S}$
corresponds to that for the \emph{reservoir} $R$ and $V^{S}$ is
the interacting potential. Using the density operator formalism to
describe the total system, the dynamical evolution, in the interaction
representation, will be given by the Liouville-Von Neumann equation,

\begin{equation}
\frac{d}{dt}\rho\left(t\right)=\frac{1}{i\hbar}\left[V\left(t\right),\rho\left(t\right)\right]\mbox{.}\label{eq:LVN}
\end{equation}

Iterating the Liouville-Von Neumann equation to second order, gives

\begin{equation}
\rho\left(t+\Delta t\right)=\rho\left(t\right)+\frac{1}{i\hbar}\int_{t}^{t+\Delta t}dt'\left[V\left(t'\right),\rho\left(t\right)\right]+\left(\frac{1}{i\hbar}\right)^{2}\int_{t}^{t+\Delta t}dt'\int_{t}^{t'}dt''\left[V\left(t'\right),\left[V\left(t''\right),\rho\left(t''\right)\right]\right]\mbox{.}
\end{equation}

Assuming that the interacting potential is sufficiently weak, the
perturbation caused in the \emph{reservoir} by the microscopic system
is negligible and we can consider that the \emph{reservoir} is in
the stationary state 
\begin{equation}
\sigma_{R}\left(t\right)\backsimeq\sigma_{R}\left(0\right)\equiv\sigma_{R}\mbox{,}
\end{equation}
where $\sigma_{R}$ is the reduced density operator for the \emph{reservoir}.
Thus, the density operator for the total system assumes the form \cite{Cohen-Tannoudji1992}

\begin{equation}
\rho\left(t\right)=\sigma_{A}\left(t\right)\otimes\sigma_{R}\left(t\right)+\rho_{cor}\left(t\right)\simeq\sigma_{A}\left(t\right)\otimes\sigma_{R}\mbox{.}
\end{equation}

The resulting master equation for the reduced density operator in
the Markov-Born approximation, or equivalently, the coarse-graining
approximation for its time evolution is

\begin{equation}
\frac{\Delta\sigma_{A}\left(t\right)}{\Delta t}=\frac{-1}{\hbar^{2}\Delta t}\int_{t}^{t+\Delta t}dt'\int_{t}^{t'}dt''Tr_{R}\left[V\left(t'\right),\left[V\left(t''\right),\sigma_{A}\left(t\right)\otimes\sigma_{R}\right]\right]\mbox{.}
\end{equation}

Assuming a separable potential

\begin{equation}
V\left(t\right)=-R\left(t\right)\otimes A\left(t\right)\mbox{,}\label{eq:potencial equa=0000E7ao mestra}
\end{equation}
and expanding the master equation commutator we can identify the function

\begin{equation}
g\left(t\right)\equiv Tr\left[\sigma_{R}R\left(t'\right)R\left(t''\right)\right]\mbox{.}\label{eq:G(t)}
\end{equation}

This is the two-point correlation function and we will use it to introduce
temperature in the theory by making the substitution

\begin{equation}
\sigma_{R}\left(t\right)\rightarrow\sigma_{R}\left(t,\beta\right)\mbox{,}
\end{equation}
so that \eqref{eq:G(t)} becomes

\begin{equation}
g\left(t,\beta\right)=\left\langle 0_{\left(\beta\right)}\right|R\left(t'\right)R\left(t''\right)\left|0_{\left(\beta\right)}\right\rangle \mbox{.}\label{eq:g.tau}
\end{equation}

Finally, we can derive the finite temperature master equation in the
coarse-graining approximation,

\begin{equation}
\begin{array}{rl}
\Delta\sigma_{A}\left(t\right) & \simeq\left(\frac{1}{i\hbar}\right)^{2}\int_{t}^{t+\Delta t}dt'\int_{t}^{t'}dt''g\left(t,\beta\right)\left(A\left(t'\right)A\left(t''\right)\sigma_{A}\left(t''\right)-A\left(t''\right)\sigma_{A}\left(t''\right)A\left(t'\right)\right)\\
\\
 & +\left(\frac{1}{i\hbar}\right)^{2}\int_{t}^{t+\Delta t}dt'\int_{t}^{t'}dt''g\left(-t,\beta\right)\left(\sigma_{A}\left(t''\right)A\left(t''\right)A\left(t'\right)-A\left(t'\right)\sigma_{A}\left(t''\right)A\left(t''\right)\right)\mbox{.}
\end{array}\label{eq:mestra.final}
\end{equation}

\subsection{The Fermion Oscillator\label{sec:The-Fermion-Oscillator}}

We consider the less common case of a Fermi oscillator and how its
population, energy levels, $\left\langle n\left|\Delta\sigma_{A}\left(t\right)\right|n\right\rangle \equiv\Delta\sigma_{A_{nn}}\left(t\right)$
behaves; let the system $A$ be a fermion oscillator of frequency
$\omega_{0}$ interacting with a \emph{reservoir} $R$, composed by
$n$ independent oscillators of frequency $\omega_{i}$ in thermal
equilibrium. So, we define the system total Hamiltonian by

\begin{equation}
H=H_{A}+H_{R}+V\mbox{,}
\end{equation}
where the Hamiltonian $H$ belongs to the Hilbert space $\mathcal{H}^{1+n}=\mathcal{H}\otimes\mathcal{H}^{n}$,
and

\begin{equation}
\begin{array}{l}
H_{A}=\hbar\omega_{0}\left(b^{S\dagger}b^{S}-\frac{1}{2}\right)\otimes\mathbb{I}\mbox{,}\\
\\
H_{R}=\mathbb{I}\otimes\hbar\underset{i}{\sum}\omega_{i}\left(a_{i}^{S\dagger}a_{i}^{S}-\frac{1}{2}\right)\mbox{,}
\end{array}
\end{equation}
where the respective ladder operators obey the usual \emph{anticommuting}
algebra, 

\begin{equation}
\begin{array}{l}
\left\{ b,b^{\dagger}\right\} =1\mbox{,}\\
\left\{ a_{i},a_{j}\right\} =\delta_{i,j}\mbox{.}
\end{array}
\end{equation}

In order to describe the interaction with the \emph{reservoir}, we
introduce the potential

\begin{equation}
V=\underset{i}{\sum}\left[g_{i}\left(\epsilon b^{S}+\eta b^{S\dagger}\right)a_{i}^{S}+g_{i}^{*}a_{i}^{S\dagger}\left(\epsilon b^{S\dagger}+\eta b^{S}\right)\right]\mbox{,}
\end{equation}
where $g_{i}$ is a coupling parameter and $\epsilon$ ,$\eta$ non-negative
constants, the expression above being the most general sesquilinear
interaction. Making $\eta=1$ and assuming the rotating wave approximation,
$\epsilon=0$, we arrive at

\begin{equation}
V=-\left(b^{S\dagger}\otimes R^{S}+b^{S}\otimes R^{S\dagger}\right)\mbox{,}
\end{equation}
where 
\begin{equation}
R^{S}=-\left(g_{1}a_{1}^{S}\otimes\mathbb{I}\otimes\mathbb{I}\cdots+\mathbb{I}\otimes g_{2}a_{2}^{S}\otimes\mathbb{I}\cdots+\cdots\right)=-\underset{i}{\sum}g_{i}a_{i}^{S}\mbox{.}
\end{equation}

Writing the master equation commutator explicitly, we have

\begin{equation}
\begin{array}{l}
Tr_{R}\left[V\left(t'\right),\left[V\left(t''\right),\sigma_{A}\left(t\right)\otimes\sigma_{R}\right]\right]\\
\\
=\left(b\left(t'\right)b^{\dagger}\left(t''\right)\sigma_{A}\left(t\right)-b^{\dagger}\left(t''\right)\sigma_{A}\left(t\right)b\left(t'\right)\right)Tr\left[\sigma_{R}R^{\dagger}\left(t'\right)R\left(t''\right)\right]\\
\\
+\left(\sigma_{A}\left(t\right)b\left(t''\right)b^{\dagger}\left(t'\right)-b^{\dagger}\left(t'\right)\sigma_{A}\left(t\right)b\left(t''\right)\right)Tr\left[\sigma_{R}R^{\dagger}\left(t''\right)R\left(t'\right)\right]\\
\\
+\left(b^{\dagger}\left(t'\right)b\left(t''\right)\sigma_{A}\left(t\right)-b\left(t''\right)\sigma_{A}\left(t\right)b^{\dagger}\left(t'\right)\right)Tr\left[\sigma_{R}R\left(t'\right)R^{\dagger}\left(t''\right)\right]\\
\\
+\left(\sigma_{A}\left(t\right)b^{\dagger}\left(t''\right)b\left(t'\right)-b\left(t'\right)\sigma_{A}\left(t\right)b^{\dagger}\left(t''\right)\right)Tr\left[\sigma_{R}R\left(t''\right)R^{\dagger}\left(t'\right)\right]\mbox{.}
\end{array}
\end{equation}

By ensuring that the vacuum of the system is separable 
\begin{equation}
\left|0_{\left(\beta\right)}\right\rangle =\left|0_{\left(\beta\right)}\right\rangle _{1}\otimes\left|0_{\left(\beta\right)}\right\rangle _{2}\otimes\cdots\mbox{,}
\end{equation}
we also guarantee the separability of the function $g\left(t,\beta\right)$,
\begin{equation}
\begin{array}{c}
Tr\left[\sigma_{R}R\left(t'\right)R\left(t''\right)\right]=\left\langle 0_{\left(\beta\right)}\right|R\left(t'\right)R\left(t''\right)\left|0_{\left(\beta\right)}\right\rangle \\
\\
=\left\langle 0_{\left(\beta\right)}\right|R\left(t'\right)R\left(t''\right)\left|0_{\left(\beta\right)}\right\rangle _{1}\left\langle 0_{\left(\beta\right)}\right|R\left(t'\right)R\left(t''\right)\left|0_{\left(\beta\right)}\right\rangle _{2}\cdots\mbox{.}
\end{array}
\end{equation}

The fermion traces follow from the above expression, in which there
appears, as expected, the Fermi-Dirac distribution, 
\begin{equation}
\begin{array}{l}
Tr\left[\sigma_{R}R^{\dagger}\left(t'\right)R\left(t''\right)\right]=\underset{i}{\sum}\left|g_{i}\right|^{2}e^{-i\omega_{i}\left(t''-t'\right)}\left(e^{\beta\hbar\omega}+1\right)^{-1}\mbox{,}\\
\\
Tr\left[\sigma_{R}R^{\dagger}\left(t''\right)R\left(t'\right)\right]=\underset{i}{\sum}\left|g_{i}\right|^{2}e^{-i\omega_{i}\left(t'-t''\right)}\left(e^{\beta\hbar\omega}+1\right)^{-1}\mbox{,}\\
\\
Tr\left[\sigma_{R}R\left(t'\right)R^{\dagger}\left(t''\right)\right]=\underset{i}{\sum}\left|g_{i}\right|^{2}e^{-i\omega_{i}\left(t'-t''\right)}\left[1-\left(e^{\beta\hbar\omega}+1\right)^{-1}\right]\mbox{,}\\
\\
Tr\left[\sigma_{R}R\left(t''\right)R^{\dagger}\left(t'\right)\right]=\underset{i}{\sum}\left|g_{i}\right|^{2}e^{-i\omega_{i}\left(t''-t'\right)}\left[1-\left(e^{\beta\hbar\omega}+1\right)^{-1}\right]\mbox{.}
\end{array}
\end{equation}

Under these considerations, we arrive at the equation for the evolution
of the ``population” of a given state $n$, for the microscopic system,

\begin{equation}
\begin{array}{c}
\frac{\Delta\sigma_{A_{nn}}\left(t\right)}{\Delta t}=-nC\sigma_{A_{nn}}\left(t\right)+\left(n+1\right)C\sigma_{A_{n+1,n+1}}\left(t\right)\\
\\
-\left(n+1\right)T_{F}\left(\sigma_{A_{n+1,n+1}}\left(t\right)+\sigma_{A_{nn}}\left(t\right)\right)+nT_{F}\left(\sigma_{A_{n-1,n-1}}\left(t\right)+\sigma_{A_{nn}}\left(t\right)\right)\mbox{,}
\end{array}\label{eq:Fermion}
\end{equation}
where 
\begin{equation}
\begin{array}{ll}
C & \equiv2\left(\frac{1}{\hbar}\right)^{2}\underset{i}{\sum}\left|g_{i}\right|^{2}\omega_{0i}^{-2}\left[\frac{1-cos\left(\omega_{0i}\Delta t\right)}{\Delta t}\right]\mbox{,}\\
\\
T_{F} & \equiv2\left(\frac{1}{\hbar}\right)^{2}\underset{i}{\sum}\left|g_{i}\right|^{2}\omega_{0i}^{-2}\left[\frac{1-cos\left(\omega_{0i}\Delta t\right)}{\Delta t}\right]\left(e^{\beta\hbar\omega_{i}}+1\right)^{-1}\mbox{.}
\end{array}
\end{equation}

Interpreting the above equation, we realize that $C$ is associated
to the spontaneous emission and $T_{F}$ to the process of stimulated
absorption and emission. More precisely, $nC$ is the spontaneous
emission rate between the states $\left|n\right\rangle $ and$\left|n+1\right\rangle $,
so the state $\left|n\right\rangle $ decays at a rate $nC$ while
the state $\left|n+1\right\rangle $ is populated at rate $\left(n+1\right)C$.
Similarly, there are processes of absorption and stimulated emission,
at rates $nT_{F}$ and $\left(n+1\right)T_{F}$, respectively. Observing
the thermodynamic limit of transition rates for low temperatures,

\[
\beta E\gg1\mbox{,}\;T_{F}\rightarrow0\mbox{,}
\]
we find that the only allowed variations are the fluctuations. However,
in the high temperatures limit, 
\[
\beta E\ll1\mbox{,}\;T_{F}\rightarrow\frac{1}{2}C\mbox{.}
\]

The transition rate takes the value $\frac{1}{2}$, due to restrictions
imposed by the fermion algebra, allowing the excitation of the only
mode of energy available. At progressively higher temperatures, the
fermion particles of the thermal bath exchange energy with the microscopic
particle through short-range correlations, exciting it or inducing
it to decay to their only two modes of oscillation, with equal probabilities.
Finally, if we want to describe a boson oscillator in a fermion heat
bath we use the expression without restricting the accessible states.

In the fermion case having only two allowed modes we can easily see
how the energy for the microscopic system evolves

\begin{equation}
\begin{array}{c}
\frac{d}{dt}\left\langle H_{A}\right\rangle _{A}=\frac{d}{dt}Tr\left[\sigma_{A}\left(t\right)H_{A}\right]=Tr\left[\left(\frac{d}{dt}\sigma_{A}\left(t\right)\right)H_{A}\right]\\
\\
=\underset{n}{\sum}\left\langle n\left|\left(\frac{d}{dt}\sigma_{A}\left(t\right)\right)H_{A}\right|n\right\rangle =\underset{n}{\sum}\hbar\omega_{0}\left(n-\frac{1}{2}\right)\left\langle n\left|\left(\frac{d}{dt}\sigma_{A}\left(t\right)\right)\right|n\right\rangle \\
\\
=-\hbar\omega_{0}\frac{1}{2}\left\langle 0\left|\left(\frac{d}{dt}\sigma_{A}\left(t\right)\right)\right|0\right\rangle +\hbar\omega_{0}\frac{1}{2}\left\langle 1\left|\left(\frac{d}{dt}\sigma_{A}\left(t\right)\right)\right|1\right\rangle \\
\\
=\hbar\omega_{0}\left(-C\sigma_{A_{1,1}}\left(t\right)+T_{F}\sigma_{A_{00}}\left(t\right)\right)\mbox{.}
\end{array}
\end{equation}

If we look for a stationary state

\begin{equation}
\frac{d}{dt}\left\langle H_{A}\right\rangle _{A}=0\mbox{,}
\end{equation}
we find that we can only reach such state with $T=0K$, so the microscopic
system energy stays oscillating in a local non equilibrium situation;
this is fundamental and enables us to construct the master equation.

\subsection{Master equation for a two-level system\label{sub:Master-2-level}}

In order to go further and integrate the equation \eqref{eq:mestra.final},
it is necessary to know the form of $\sigma_{A}\left(t''\right)$.
For this reason, we propose a two-level atomic system described by
the Hamiltonian

\begin{equation}
H_{A}=\omega_{0}R_{3}\left(\tau\right)\mbox{,}
\end{equation}
where $R_{3}=\frac{1}{2}\left|+\right\rangle \left\langle +\right|-\frac{1}{2}\left|-\right\rangle \left\langle -\right|$
and $\tau$ is the proper time of the atom, whose eigenvalues ​​are

\begin{equation}
\begin{array}{l}
H_{A}\left|+\right\rangle =\frac{1}{2}\omega_{0}\left|+\right\rangle \mbox{,}\\
\\
H_{A}\left|-\right\rangle =-\frac{1}{2}\omega_{0}\left|-\right\rangle \mbox{.}
\end{array}
\end{equation}

The proposed field-atom interaction is similar to \eqref{eq:Hamiltoniano.intera=0000E7ao.DeWitt}
and given by the Hamiltonian

\begin{equation}
\begin{array}{c}
H_{I}=\mu R_{2}\left(\tau\right)\phi\left(\tau\right)\mbox{,}\end{array}\label{eq:hamiltoniano.intera=0000E7ao.audretsch}
\end{equation}
where $\mu$ is the coupling constant and $R_{2}=\frac{i}{2}\left(R_{-}-R_{+}\right)$,
with $R_{+}=\frac{i}{2}\left|+\right\rangle \left\langle -\right|$
and $R_{-}=\frac{i}{2}\left|-\right\rangle \left\langle +\right|$.
It is noteworthy that the operators $R_{+}$ and $R_{-}$ can be seen
as raising and lowering operators of the atomic system and, together
with $R_{3}$, generate the $SO(3)$ algebra,

\begin{equation}
\left[R_{+},R_{-}\right]=2R_{3},\label{eq:rela=0000E7ao.comuta=0000E7ao.R+.R-}
\end{equation}

\begin{equation}
\left[R_{3},R_{\pm}\right]=\pm R_{\pm}\mbox{.}\label{eq:rela=0000E7ao.comuta=0000E7ao.R3.R+-}
\end{equation}

From the Heisenberg equation 
\begin{equation}
i\frac{d}{dt}A\left(t\right)=\left[A\left(t\right),H_{T}\right]\mbox{,}\label{eq:eq.heisenberg}
\end{equation}
where $H_{T}=H_{A}+H_{F}+H_{I}$, we obtain the evolution equations

\begin{equation}
\frac{d}{d\tau}R_{\pm}\left(\tau\right)=\pm i\omega_{0}R_{\pm}\left(\tau\right)+i\mu\phi\left(\tau\right)\left[R_{2}\left(\tau\right),R_{\pm}\left(\tau\right)\right]\mbox{,}\label{eq:eq.mov.r+}
\end{equation}

\begin{equation}
\frac{d}{d\tau}R_{3}\left(\tau\right)=i\mu\phi\left(\tau\right)\left[R_{2}\left(\tau\right),R_{3}\left(\tau\right)\right]\mbox{.}\label{eq:eq.mov.R3}
\end{equation}

Working on \eqref{eq:eq.mov.r+} and \eqref{eq:eq.mov.R3} to first
order in $\mu$ we can conveniently breaking then down into a free
part, which preserves in the absence of coupling, 

\begin{equation}
R_{\pm}^{f}\left(\tau\right)=R_{\pm}^{f}\left(\tau_{0}\right)e^{\pm i\omega_{0}\left(\tau-\tau_{0}\right)}\mbox{,}\label{eq:R+-.livre}
\end{equation}

\begin{equation}
R_{3}^{f}\left(\tau\right)=R_{3}\left(\tau_{0}\right)\mbox{,}
\end{equation}
and another part of the resulting coupling

\begin{equation}
R_{\pm}^{S}\left(\tau\right)=i\mu\intop_{\tau_{0}}^{\tau}d\tau'\phi^{f}\left(\tau'\right)\left[R_{2}^{f}\left(\tau'\right),R_{\pm}^{f}\left(\tau\right)\right]\mbox{,}\label{eq:R+.source}
\end{equation}

\begin{equation}
R_{3}^{S}\left(\tau\right)=i\mu\intop_{\tau_{0}}^{\tau}d\tau'\phi^{f}\left(\tau'\right)\left[R_{2}^{f}\left(\tau'\right),R_{3}^{f}\left(\tau\right)\right]\mbox{,}\label{eq:R3.source}
\end{equation}
where $\phi^{f}\left(\tau\right)$ is the free part of the field,
$\phi\left(\tau\right)=\phi^{f}\left(\tau\right)+\phi^{S}\left(\tau\right)$,

\begin{equation}
\phi^{f}\left(\tau\right)=\left(2\pi\right)^{-3/2}\int\frac{d^{3}\mathrm{k}}{\sqrt{2\omega_{\mathbf{k}}}}\left[a_{\mathbf{k}}e^{-i\left(\omega_{\mathbf{k}}t-\mathbf{k}.\mathbf{x}\right)}+a_{\mathbf{k}}^{\dagger}e^{i\left(\omega_{\mathbf{k}}t-\mathbf{k}.\mathbf{x}\right)}\right]\mbox{,}\label{eq:eq.campo.part.livre}
\end{equation}

\begin{equation}
\phi^{S}\left(\tau\right)=i\mu\intop_{\tau_{0}}^{\tau}d\tau'R_{2}^{f}\left(\tau'\right)\left[\phi^{f}\left(\tau'\right),\phi^{f}\left(\tau\right)\right]\mbox{.}\label{eq:eq.campo.part.intera=0000E7ao}
\end{equation}

Therefore, we write the atom energy variation, $H_{A}\left(\tau\right)$,
as
\begin{equation}
\frac{d}{d\tau}H_{A}\left(\tau\right)=-\mu\phi\left(\tau\right)\frac{d}{d\tau}R_{2}^{f}\left(\tau\right)-i\mu^{2}\intop_{\tau_{0}}^{\tau}d\tau'\phi\left(\tau\right)\phi^{f}\left(\tau'\right)\frac{d}{d\tau}\left[R_{2}^{f}\left(\tau'\right),R_{2}^{f}\left(\tau\right)\right]\mbox{.}\label{eq:evolu=0000E7ao.H.A}
\end{equation}

One of the purpose of this section is to identify and quantitatively
analyze the contributions from the vacuum fluctuations and radiation
reaction, so, we can be identified that the free part of the filed
\eqref{eq:eq.campo.part.livre} is related to the vacuum fluctuations
and, consequently, the radiation reaction is caused by the interaction
part \eqref{eq:eq.campo.part.intera=0000E7ao}. The energy is a physical
observable and, as such, represented by a hermitian operator, we would
like to recognize these two different contributions on the energy
variation and associates them with  two physical processes which are
represented by hermitian operators. For this condition to be valid
it is necessary to perform the ordering of operators in \eqref{eq:evolu=0000E7ao.H.A}
. Acknowledging that in quantum theory exists one indeterminacy as
to the order of operators, lets consider the product $\lambda AB+\left(1-\lambda\right)BA$ ,
resulting for the vacuum fluctuation

\begin{equation}
\begin{array}{c}
\left(\frac{d}{d\tau}H_{A}\left(\tau\right)\right)_{VF}=\lambda\left\{ -\mu\phi^{f}\left(\tau\right)\frac{d}{d\tau}R_{2}^{f}\left(\tau\right)-i\mu^{2}\intop_{\tau_{0}}^{\tau}d\tau'\phi^{f}\left(\tau\right)\phi^{f}\left(\tau'\right)\frac{d}{d\tau}\left[R_{2}^{f}\left(\tau'\right),R_{2}^{f}\left(\tau\right)\right]\right\} \\
\\
+\left(1-\lambda\right)\left\{ -\mu\frac{d}{d\tau}R_{2}^{f}\left(\tau\right)\phi^{f}\left(\tau\right)-i\mu^{2}\intop_{\tau_{0}}^{\tau}d\tau'\phi^{f}\left(\tau'\right)\phi^{f}\left(\tau\right)\frac{d}{d\tau}\left[R_{2}^{f}\left(\tau'\right),R_{2}^{f}\left(\tau\right)\right]\right\} \mbox{,}
\end{array}
\end{equation}
and for the radiation reaction 

\begin{equation}
\left(\frac{d}{d\tau}H_{A}\left(\tau\right)\right)_{RR}=-i\mu^{2}\intop_{\tau_{0}}^{\tau}d\tau'\left[\phi^{f}\left(\tau'\right),\phi^{f}\left(\tau\right)\right]\left[\lambda R_{2}^{f}\left(\tau'\right)\frac{d}{d\tau}R_{2}^{f}\left(\tau\right)+\left(1-\lambda\right)\left(\frac{d}{d\tau}R_{2}^{f}\left(\tau\right)\right)R_{2}^{f}\left(\tau'\right)\right]\mbox{.}
\end{equation}

Taking the expected value $\left\langle \frac{d}{d\tau}H_{A}\left(\tau\right)\right\rangle \equiv\left\langle a\left|\otimes\left\langle 0\left|\frac{d}{d\tau}H_{A}\left(\tau\right)\right|0\right\rangle \otimes\right|a\right\rangle $,
where $\left|0\right\rangle $ is the vacuum field and $\left|a\right\rangle $
an arbitrary atom state, results 

\begin{equation}
\begin{array}{c}
\left\langle \frac{d}{d\tau}H_{A}\left(\tau\right)\right\rangle _{VF}=i\mu^{2}\intop_{\tau_{0}}^{\tau}d\tau'\frac{d}{d\tau}\left\langle a\left|\left[R_{2}^{f}\left(\tau\right),R_{2}^{f}\left(\tau'\right)\right]\right|a\right\rangle \times\\
\\
\left\langle 0\left|\left(\lambda-\frac{1}{2}\right)\left[\phi^{f}\left(\tau\right),\phi^{f}\left(\tau'\right)\right]+\frac{1}{2}\left[\phi^{f}\left(\tau\right),\phi^{f}\left(\tau'\right)\right]_{+}\right|0\right\rangle \mbox{,}
\end{array}\label{eq:VF.valor.esperado}
\end{equation}

$ $

\begin{equation}
\begin{array}{c}
\left\langle \frac{d}{d\tau}H_{A}\left(\tau\right)\right\rangle _{RR}=i\mu^{2}\intop_{\tau_{0}}^{\tau}d\tau'\left\langle 0\left|\left[\phi^{f}\left(\tau\right),\phi^{f}\left(\tau'\right)\right]\right|0\right\rangle \times\\
\\
\frac{d}{d\tau}\left\langle a\left|\frac{1}{2}\left[R_{2}^{f}\left(\tau\right),R_{2}^{f}\left(\tau'\right)\right]_{+}-\left(\lambda-\frac{1}{2}\right)\left[R_{2}^{f}\left(\tau\right),R_{2}^{f}\left(\tau'\right)\right]\right|a\right\rangle \mbox{.}
\end{array}\label{eq:RR.valor.esperado}
\end{equation}

In equations \eqref{eq:VF.valor.esperado} and \eqref{eq:RR.valor.esperado}
we can recognize the commonly known functions in quantum optics as
correlation function $C\left(\tau,\tau'\right)$ and linear susceptibility
$\chi\left(\tau,\tau'\right)$ . For the field, the correlation and
susceptibility functions are

\begin{equation}
C^{F}\left(\tau,\tau'\right)=\frac{1}{2}\left\langle 0\left|\left[\phi^{f}\left(\tau\right),\phi^{f}\left(\tau'\right)\right]_{+}\right|0\right\rangle \mbox{,}\label{eq:func.correl.campo}
\end{equation}

\begin{equation}
\chi^{F}\left(\tau,\tau'\right)=\frac{1}{2i}\left\langle 0\left|\left[\phi^{f}\left(\tau\right),\phi^{f}\left(\tau'\right)\right]\right|0\right\rangle \mbox{,}\label{eq:func.suscep.campo}
\end{equation}
and for the atomic system

\begin{equation}
C^{A}\left(\tau,\tau'\right)=\frac{1}{2}\left\langle a\left|\left[R_{2}^{f}\left(\tau\right),R_{2}^{f}\left(\tau'\right)\right]_{+}\right|a\right\rangle \mbox{,}\label{eq:func.correl.atomo}
\end{equation}

\begin{equation}
\chi^{A}\left(\tau,\tau'\right)=\frac{1}{2i}\left\langle a\left|\left[R_{2}^{f}\left(\tau\right),R_{2}^{f}\left(\tau'\right)\right]\right|a\right\rangle \mbox{.}\label{eq:func.susep.atomo}
\end{equation}

Explicit calculation of the correlation and susceptibility functions
gives, to the atomic system, 

\begin{equation}
\begin{array}{c}
C^{A}\left(\tau,\tau'\right)=\frac{1}{4}\,cos\left(\omega_{0}\left(\tau-\tau'\right)\right)\mbox{,}\end{array}
\end{equation}

\begin{equation}
\begin{array}{c}
\chi^{A}\left(\tau,\tau'\right)=\frac{1}{2}\,sin\left(\omega_{0}\left(\tau-\tau'\right)\right)\left\langle a\left|R_{3}^{f}\left(\tau_{0}\right)\right|a\right\rangle \mbox{,}\end{array}
\end{equation}
and, using Wightman functions, to the field

\begin{equation}
C^{F}\left(\tau,\tau'\right)=\frac{1}{8\pi^{2}}\left[\frac{1}{\left[\left|\Delta\mathbf{x}\right|^{2}-\left(\Delta t-i\epsilon\right)^{2}\right]}+\frac{1}{\left[\left|\Delta\mathbf{x}\right|^{2}-\left(\Delta t+i\epsilon\right)^{2}\right]}\right]\mbox{,}\label{eq:func.correl.campo.explicita}
\end{equation}

\begin{equation}
\chi^{F}\left(\tau,\tau'\right)=\frac{1}{8\pi^{2}i}\left[\frac{1}{\left[\left|\Delta\mathbf{x}\right|^{2}-\left(\Delta t-i\epsilon\right)^{2}\right]}-\frac{1}{\left[\left|\Delta\mathbf{x}\right|^{2}-\left(\Delta t+i\epsilon\right)^{2}\right]}\right].\label{eq:func.suscep.campo.explicita}
\end{equation}

To go beyond \eqref{eq:func.correl.campo.explicita} and \eqref{eq:func.suscep.campo.explicita}
one must choose a parametrization, thus, using the uniformly accelerated
case as before \eqref{eq:param.acelerada} results

\begin{equation}
C^{F}\left(\tau,\tau'\right)=-\frac{1}{8\pi^{2}}\underset{n=-\infty}{\overset{\infty}{\sum}}\left[\frac{1}{\left(\Delta\mathbf{\tau}-i\epsilon2+i\frac{2\pi}{\alpha}n\right)^{2}}+\frac{1}{\left(\Delta\mathbf{\tau}+i\epsilon2+i\frac{2\pi}{\alpha}n\right)^{2}}\right]\mbox{,}\label{eq:func.correl.campo.explicita.inercial-1}
\end{equation}

\begin{equation}
\chi^{F}\left(\tau,\tau'\right)=-\frac{1}{8\pi^{2}i}\underset{n=-\infty}{\overset{\infty}{\sum}}\left[\frac{1}{\left(\Delta\mathbf{\tau}-i\epsilon2+i\frac{2\pi}{\alpha}n\right)^{2}}-\frac{1}{\left(\Delta\mathbf{\tau}+i\epsilon2+i\frac{2\pi}{\alpha}n\right)^{2}}\right]\mbox{.}\label{eq:func.suscep.campo.explicita.inercia-1}
\end{equation}

Unfortunately, in this case, the integrals found in \eqref{eq:VF.valor.esperado}
and \eqref{eq:RR.valor.esperado} are divergent. One option to overcome
this situation is to restrict the study to the asymptotic case, $\tau_{0}\rightarrow-\infty$,
leading to

\begin{equation}
\left\langle \frac{d}{d\tau}H_{A}\left(\tau\right)\right\rangle _{VF}=-\omega_{0}\mu^{2}\left\langle a\left|R_{3}\right|a\right\rangle \left[\frac{\omega_{0}}{8\pi}\left(1+\frac{2}{e^{\frac{2\pi\omega_{0}}{\alpha}}-1}\right)+i\left(2\lambda-1\right)\intop_{0}^{\infty}du\,cos\left(\omega_{0}u\right)\chi^{F}\left(u\right)\right]\mbox{,}\label{eq:VFi.final}
\end{equation}

\begin{equation}
\left\langle \frac{d}{d\tau}H_{A}\left(\tau\right)\right\rangle _{RR}=\omega_{0}\mu^{2}\left[-\frac{\omega_{0}}{16\pi}\left(1+\frac{2}{e^{\frac{2\pi\omega_{0}}{\alpha}}-1}\right)+i\left(2\lambda-1\right)\left\langle a\left|R_{3}\right|a\right\rangle \intop_{0}^{\infty}du\,cos\left(\omega_{0}u\right)\chi^{F}\left(u\right)\right]\mbox{,}\label{eq:RR.final}
\end{equation}
where $u=\mathbf{\tau}-\mathbf{\tau}'$. Nevertheless, the integral

\begin{equation}
\intop_{0}^{\infty}du\,cos\left(\omega_{0}u\right)\chi^{F}\left(u\right)=-\frac{1}{8\pi^{2}i}\underset{n=-\infty}{\overset{\infty}{\sum}}\intop_{0}^{\infty}du\,cos\left(\omega_{0}u\right)\left[\frac{1}{\left(u-i\epsilon2+i\frac{2\pi}{\alpha}n\right)^{2}}-\frac{1}{\left(u+i\epsilon2+i\frac{2\pi}{\alpha}n\right)^{2}}\right],
\end{equation}
is still diverging. From equations \eqref{eq:VFi.final} and \eqref{eq:RR.final}
can be concluded two important facts. For the physical observable
associated with vacuum fluctuations and radiation reaction be independent
and represented by finite quantities it is necessary to perform operators
ordering. It's also clear that the only possible order to yield not
divergent rates is the symmetrical one, $\lambda=\frac{1}{2}$. Thus,
for the symmetric case,

\begin{equation}
\left\langle \frac{d}{d\tau}H_{A}\left(\tau\right)\right\rangle _{VF}=-\frac{\omega_{0}^{2}\mu^{2}}{8\pi}\left\langle a\left|R_{3}\right|a\right\rangle \left[1+\frac{2}{e^{\frac{2\pi\omega_{0}}{\alpha}}-1}\right]\mbox{,}\label{eq:VF.rindleri.final.simetrico}
\end{equation}

\begin{equation}
\left\langle \frac{d}{d\tau}H_{A}\left(\tau\right)\right\rangle _{RR}=-\frac{\omega_{0}^{2}\mu^{2}}{16\pi}\left[1+\frac{2}{e^{\frac{2\pi\omega_{0}}{\alpha}}-1}\right]\mbox{.}\label{eq:RR.rindler.final.simetrico}
\end{equation}

From \eqref{eq:RR.rindler.final.simetrico}, we notice that the radiation
reaction is responsible for giving energy to the field and has a purely
dissipative nature, independent of $\mu$ and $\omega_{0}$. To better
interpret the vacuum fluctuations \eqref{eq:VF.rindleri.final.simetrico},
we replace $R_{3}$ for its explicit form

\begin{equation}
\left\langle \frac{d}{d\tau}H_{A}\left(\tau\right)\right\rangle _{VF}=-\frac{\omega_{0}^{2}\mu^{2}}{16\pi}\left[1+\frac{2}{e^{\frac{2\pi\omega_{0}}{\alpha}}-1}\right]\left[\left|\left.\left\langle a\right|+\right\rangle \right|^{2}-\left|\left.\left\langle a\right|-\right\rangle \right|^{2}\right]\mbox{,}
\end{equation}
therefore, in the case where $\left|a\right\rangle =\left|+\right\rangle $,
i.e. the atom is in the excited state, results 
\begin{equation}
\left\langle +\left|\frac{d}{d\tau}H_{A}\left(\tau\right)\right|+\right\rangle _{VF}=-\frac{\omega_{0}^{2}\mu^{2}}{16\pi}\left[1+\frac{2}{e^{\frac{2\pi\omega_{0}}{\alpha}}-1}\right]\mbox{,}
\end{equation}
and, as expected, the vacuum fluctuation is responsible for giving
energy. In case that $\left|a\right\rangle =\left|-\right\rangle $,
which is equivalent to the ground state,

\begin{equation}
\left\langle -\left|\frac{d}{d\tau}H_{A}\left(\tau\right)\right|-\right\rangle _{VF}=\frac{\omega_{0}^{2}\mu^{2}}{16\pi}\left[1+\frac{2}{e^{\frac{2\pi\omega_{0}}{\alpha}}-1}\right]\mbox{,}
\end{equation}
so the vacuum fluctuation is responsible for excite the atom. Note
that, on average, vacuum fluctuations cause a zero energy change,
as expected, since it is related to the free part of the field and
so taking place regardless of the interaction. Therefore, it is not
unusual to expect that the average variation caused by it to be zero. 

Returning to equations \eqref{eq:VFi.final} and \eqref{eq:RR.final}
, observe that when adding both

\begin{equation}
\left\langle \frac{d}{d\tau}H_{A}\left(\tau\right)\right\rangle _{T}=-\frac{\omega_{0}^{2}\mu^{2}}{8\pi}\left[\frac{1}{2}+\left\langle a\left|R_{3}\right|a\right\rangle \right]\left[1+\frac{2}{e^{\frac{2\pi\omega_{0}}{\alpha}}-1}\right]\mbox{,}\label{eq:flutua=0000E7ao.total.final}
\end{equation}
and the equation \eqref{eq:flutua=0000E7ao.total.final} has no dependence
of the parameter $\lambda$,  that is, for the total physical process
any ordering results in the same energy variation \cite{Tomazelli2003}.

Keeping this in mind, we now establish the following correspondences
between \eqref{eq:hamiltoniano.intera=0000E7ao.audretsch} and \eqref{eq:potencial equa=0000E7ao mestra}
:

\begin{equation}
\left\{ \begin{array}{l}
t\rightarrow\tau\,\mbox{,}\\
A\left(t\right)\rightarrow R_{2}\left(\tau\right)\,\mbox{,}\\
R\left(t\right)\rightarrow\phi\left(\tau\right)\,\mbox{.}
\end{array}\right.
\end{equation}

Here, equation \eqref{eq:mestra.final} is projected in the basis
of the eigenstates of the atomic system. Thus, taking the expectation
value of equation \eqref{eq:mestra.final} in an arbitrary state $\left|a\right\rangle $
of the atomic system and defined the contracted notation $\Delta\sigma_{A_{aa}}\left(\tau\right)\equiv\left\langle a\left|\Delta\sigma_{A}\left(\tau\right)\right|a\right\rangle \mbox{,}$
equation becomes

\begin{equation}
\frac{\Delta\sigma_{A_{aa}}\left(\tau\right)}{\Delta\tau}\simeq-\frac{1}{4\Delta\tau}\int_{\tau}^{\tau+\Delta\tau}d\tau'\int_{\tau}^{\tau'}d\tau''g\left(\tau,\beta\right)\left[e^{ai\omega_{0}\left(\tau'-\tau''\right)}\sigma_{A_{aa}}\left(\tau\right)-e^{-ai\omega_{0}\left(\tau'-\tau''\right)}\sigma_{A_{bb}}\left(\tau\right)\right]+h.c.\,\mbox{,}\label{eq:eq.mestra.2.niveis}
\end{equation}
where $h.c.$ represents the hermitian conjugate. Equation \eqref{eq:g.tau}
is, in this case, $g\left(\tau,\beta\right)=\left\langle 0_{\left(\beta\right)}\right|\phi\left(\tau'\right)\phi\left(\tau''\right)\left|0_{\left(\beta\right)}\right\rangle $.
Replacing the fields, we obtain

\[
g\left(\tau',\tau''\right)=\left(2\pi\right)^{-3}\int\frac{d^{3}\mathrm{k}}{2\omega_{\mathbf{k}}}\left[\left\langle 0_{\left(\beta\right)}\right|N_{\mathbf{k}}\left|0_{\left(\beta\right)}\right\rangle e^{i\omega_{\mathbf{\mathbf{k}}}\left(\tau'-\tau''\right)-i\mathbf{k}.\left(\mathbf{x'-x''}\right)}+\left(\left\langle 0_{\left(\beta\right)}\right|N_{\mathbf{k}}\left|0_{\left(\beta\right)}\right\rangle +1\right)e^{-i\omega_{\mathbf{\mathbf{k}}}\left(\tau'-\tau''\right)+i\mathbf{k}.\left(\mathbf{x'-x''}\right)}\right]\mbox{,}
\]
or, using the results found for the expected value of the number operator
\eqref{eq:est.bos.einstein}, 

\begin{equation}
g\left(\tau',\tau''\right)=\left(2\pi\right)^{-3}\int\frac{d^{3}\mathrm{k}'}{2\omega_{\mathbf{\mathbf{k}}}}\left[\left[e^{\beta\omega_{\mathbf{k}}}-1\right]^{-1}e^{i\omega_{\mathbf{\mathbf{k}}}\left(\tau'-\tau''\right)-i\mathbf{k}.\left(\mathbf{x'-x''}\right)}+\left[1-e^{-\beta\omega_{\mathbf{k}}}\right]^{-1}e^{-i\omega_{\mathbf{\mathbf{k}}}\left(\tau'-\tau''\right)+i\mathbf{k}.\left(\mathbf{x'-x''}\right)}\right]\mbox{.}\label{eq:g.tau.temperatura}
\end{equation}

Next, consider the master equation in the inertial case and, subsequently,
in the case of an accelerated frame. In the inertial case, 

\begin{equation}
\begin{array}{c}
g\left(\tau',\tau''\right)=\frac{1}{4\pi^{2}}\underset{n=-\infty}{\overset{\infty}{\sum}}\frac{1}{i2\beta\left(\gamma-1\right)\left(\tau'-\tau''\right)\left(n+\epsilon\right)-\left[\left(\tau'-\tau''\right)-i\beta\left(n+\epsilon\right)\right]^{2}}\\
\\
=\frac{\sqrt{1-v^{2}}\left[coth\left(\frac{(v-1)}{\sqrt{1-v^{2}}}\frac{\pi}{\beta}\left(\tau'-\tau''\right)\right)+coth\left(\frac{(v+1)}{\sqrt{1-v^{2}}}\frac{\pi}{\beta}\left(\tau'-\tau''\right)\right)\right]}{8\pi\beta v\left(\tau'-\tau''\right)}\mbox{.}
\end{array}
\end{equation}

The above expression is too complicated to be integrated. However,
in the limit $v\rightarrow0$, 

\begin{equation}
g\left(\tau',\tau''\right)=-\frac{1}{4\pi^{2}}\underset{n=-\infty}{\overset{\infty}{\sum}}\frac{1}{\left[\left(\tau'-\tau''\right)-i\beta\left(n+\epsilon\right)\right]^{2}}=-\frac{1}{4\beta^{2}}csch\left(\frac{\pi\left(\tau'-\tau''\right)}{\beta}\right)^{2}\mbox{.}
\end{equation}

In this case, equation \eqref{eq:eq.mestra.2.niveis} becomes

\begin{equation}
\frac{\Delta\sigma_{A_{aa}}\left(\tau\right)}{\Delta\tau}\simeq\frac{1}{16\pi^{2}}\underset{n=-\infty}{\overset{\infty}{\sum}}\int_{-\infty}^{\infty}du\,\left[u-i\beta\left(n+\epsilon\right)\right]^{-2}\left[e^{ai\omega_{0}u}\sigma_{A_{aa}}\left(\tau\right)-e^{-ai\omega_{0}u}\sigma_{A_{bb}}\left(\tau\right)\right]\mbox{.}\label{eq:eq.mestra.integral.minkoski-1}
\end{equation}

Writing the master equation \eqref{eq:eq.mestra.integral.minkoski-1}
for the states $\left|+\right\rangle $ e $\left|-\right\rangle $,
results

\begin{equation}
\frac{\Delta\sigma_{+}\left(\tau\right)}{\Delta\tau}\simeq-\frac{\omega_{0}}{8\pi}\left\{ \sigma_{-}\left(\tau\right)+\frac{1}{1-e^{-\omega_{0}\beta}}\left[\sigma_{+}\left(\tau\right)-\sigma_{-}\left(\tau\right)\right]\right\} ,\label{eq:delta.+}
\end{equation}

\begin{equation}
\frac{\Delta\sigma_{-}\left(\tau\right)}{\Delta\tau}\simeq\frac{\omega_{0}}{8\pi}\left\{ \sigma_{-}\left(\tau\right)+\frac{1}{1-e^{-\omega_{0}\beta}}\left[\sigma_{+}\left(\tau\right)-\sigma_{-}\left(\tau\right)\right]\right\} \mbox{.}\label{eq:delta.-}
\end{equation}

From \eqref{eq:delta.+} e \eqref{eq:delta.-}, it follows that

\begin{equation}
\frac{\Delta\sigma_{+}\left(\tau\right)}{\Delta\tau}=-\frac{\Delta\sigma_{-}\left(\tau\right)}{\Delta\tau}\mbox{.}
\end{equation}
At steady regime, $\frac{\Delta\sigma_{+}\left(\tau\right)}{\Delta\tau}=0=\frac{\Delta\sigma_{-}\left(\tau\right)}{\Delta\tau}$,
we find the detailed balance condition

\begin{equation}
\frac{\sigma_{+}\left(\tau\right)}{\sigma_{-}\left(\tau\right)}=e^{-\omega_{0}\beta}\mbox{.}
\end{equation}

Solving the coupled equations \eqref{eq:delta.+} and \eqref{eq:delta.-}
, with the initial conditions$\sigma_{+}\left(0\right)=\sigma_{0+}$
e $\sigma_{-}\left(0\right)=\sigma_{0-}$and the boundary conditions
$\sigma_{+}\left(\tau\right)+\sigma_{-}\left(\tau\right)=1$, we find

\begin{equation}
\sigma_{+}\left(\tau\right)=\sigma_{0+}e^{-coth\left(\frac{1}{2}\omega_{0}\beta\right)\frac{\omega_{0}}{8\pi}\tau}-\frac{e^{-coth\left(\frac{1}{2}\omega_{0}\beta\right)\frac{\omega_{0}}{8\pi}\tau}}{1+e^{\omega_{0}\beta}}\left[1-e^{coth\left(\frac{1}{2}\omega_{0}\beta\right)\frac{\omega_{0}}{8\pi}\tau}\right]\mbox{,}\label{eq:evolu=0000E7ao.eq.mestra.minkoswki.sigma+.cte}
\end{equation}

\begin{equation}
\sigma_{-}\left(\tau\right)=-\sigma_{0+}e^{-coth\left(\frac{1}{2}\omega_{0}\beta\right)\frac{\omega_{0}}{8\pi}\tau}+\frac{e^{-coth\left(\frac{1}{2}\omega_{0}\beta\right)\frac{\omega_{0}}{8\pi}\tau}}{1+e^{\omega_{0}\beta}}\left[1+e^{\omega_{o}\beta}e^{coth\left(\frac{1}{2}\omega_{0}\beta\right)\frac{\omega_{0}}{8\pi}\tau}\right]\mbox{.}\label{eq:evolu=0000E7ao.eq.mestra.minkoswki.sigma-.cte}
\end{equation}

\begin{figure}
\centering{}\includegraphics[width=8cm]{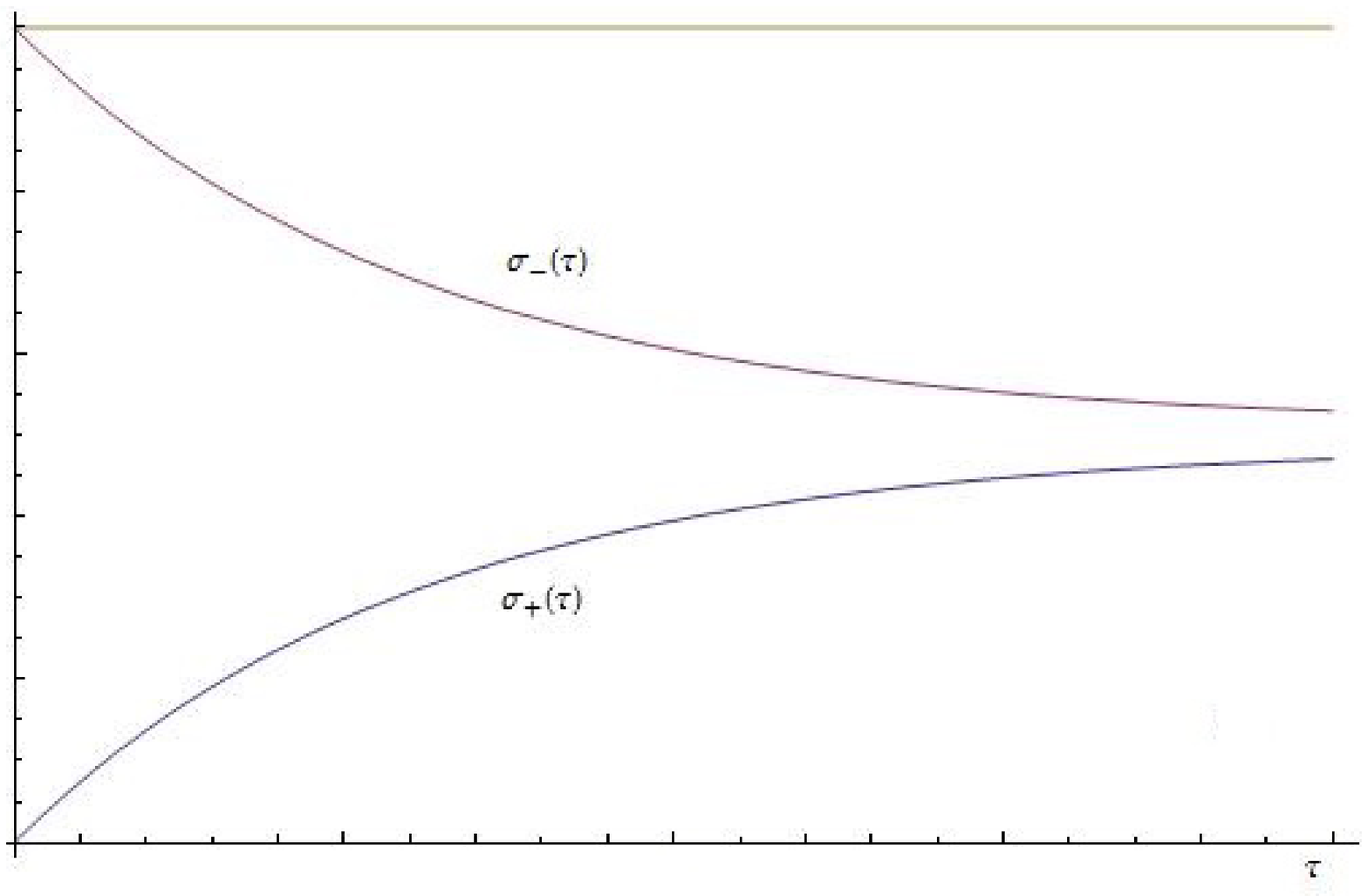}\includegraphics[width=8cm]{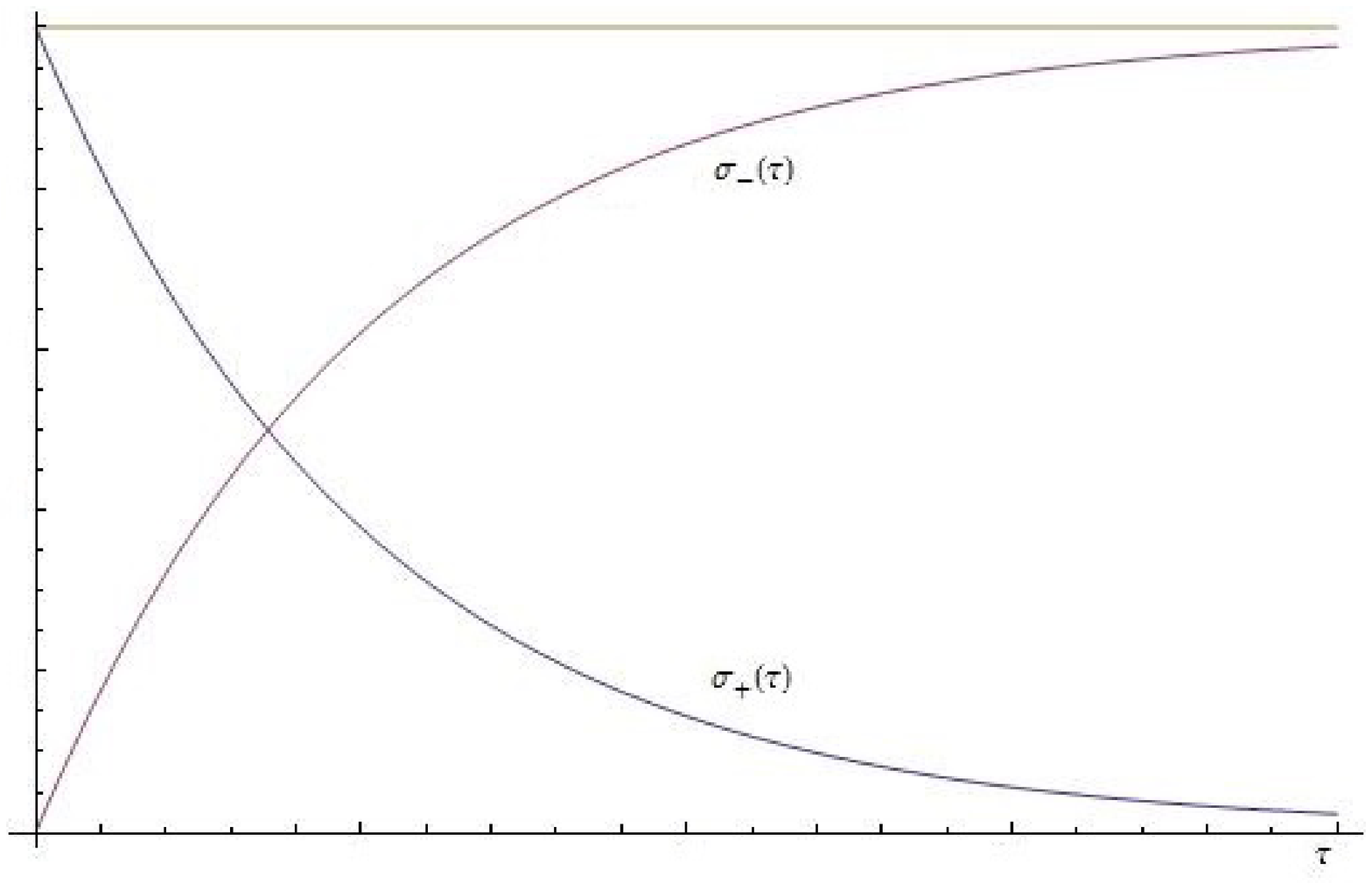}\protect\caption{The evolution of $\sigma_{+}\left(\tau\right)$ and $\sigma_{-}\left(\tau\right)$
at high (left) and low (right) temperature regimes.}
\label{eq.mestra.minkowski}
\end{figure}

The above equations \eqref{eq:evolu=0000E7ao.eq.mestra.minkoswki.sigma+.cte}
and \eqref{eq:evolu=0000E7ao.eq.mestra.minkoswki.sigma-.cte} exhibit
a non-trivial intertwining among $\sigma_{0+}$, $\beta$, $\omega_{0}$
that determines the time evolution of the system. Figure \ref{eq.mestra.minkowski}
depicts two distinct situations. The left curves correspond to the
high temperature behaviour of the system, $\omega_{0}\beta\ll1$;
in this case, the system evolves from a state in which $\sigma_{+}\left(0\right)\ll\sigma_{-}\left(0\right)$,
reaching an equilibrium state where both levels are equally filled.
At low temperatures, as illustrated by the graph at right for $\omega_{0}\beta\gg1$,
when there are initially many excited particles, $\sigma_{+}\left(0\right)\gg\sigma_{-}\left(0\right)$,
the system evolves in such a way to completely fill $\sigma_{-}$.
The intermediate conditions vary between these two extremes. Also,
it turns out that the $\beta$ is primarily responsible for determining
the values of $\sigma_{\pm}\left(\infty\right)$ for which the system
converges, the parameter $\omega_{0}$ giving the rate in which it
converges.

Finally, in the limit $\tau\rightarrow\infty$, we obtain in both
situations

\[
\sigma_{+}\left(\infty\right)=\frac{1}{1+e^{\omega_{0}\beta}}\mbox{,}
\]

\[
\sigma_{-}\left(\infty\right)=\frac{e^{\omega_{0}\beta}}{1+e^{\omega_{0}\beta}}\mbox{.}
\]

We then conclude that the two-level atomic system obeys the Fermi-Dirac
statistics, as expected, according to Bloch theorem. 

Considering now the uniformly accelerated frame with parametrization
\eqref{eq:param.acelerada}, we obtain

\begin{equation}
\begin{array}{c}
g\left(\tau',\tau''\right)=\frac{\alpha^{2}}{4\pi^{2}}\underset{n=-\infty}{\overset{\infty}{\sum}}\dfrac{1}{\left|cosh\left(\alpha\tau'\right)-cosh\left(\alpha\tau''\right)\right|^{2}-\left[sinh\left(\alpha\tau'\right)-sinh\left(\alpha\tau''\right)-i\beta\alpha n\right]^{2}}\\
\\
\\
=\dfrac{\alpha\left[coth\left(\frac{\pi\left(e^{\alpha\tau'}-e^{\alpha\tau''}\right)}{\alpha\beta}\right)-coth\left(\frac{2\pi e^{-\frac{1}{2}\alpha\left(\tau'+\tau''\right)}sinh\left(\frac{1}{2}\alpha\left(\tau'-\tau''\right)\right)}{\alpha\beta}\right)\right]}{8\pi\beta\left[cosh\left(\alpha\tau'\right)-cosh\left(\alpha\tau''\right)\right]}\mbox{.}
\end{array}\label{eq:eq.mestra.integral.Rindler}
\end{equation}

As in the inertial case, the equation \eqref{eq:eq.mestra.integral.Rindler}
is too complex to be integrated. However, limiting cases provide us
remarkable results. Thus,in the limit $\alpha\rightarrow0$, we obtains

\begin{equation}
g\left(\tau',\tau''\right)=-\frac{1}{4\beta^{2}}csch\left(\frac{\pi\left(\tau'-\tau''\right)}{\beta}\right)^{2}\mbox{,}\label{eq:g.rindler.acelera=0000E7ao.zero}
\end{equation}
recovering, as expected, the inertial result and assuring the correctness
of $g\left(\tau',\tau''\right)$. In the limit of low temperatures,
ie, $\beta\rightarrow\infty$  , we find

\begin{equation}
g\left(\tau',\tau''\right)=-\frac{\alpha^{2}}{16\pi^{2}}csch\left(\frac{\alpha\left(\tau'-\tau''\right)}{2}\right)^{2}\mbox{.}\label{eq:g.rindler.temperatura.zero}
\end{equation}

By comparing \eqref{eq:g.rindler.acelera=0000E7ao.zero} with \eqref{eq:g.rindler.temperatura.zero},
we obtain the relation proposed by Unruh,

\[
T=\frac{\alpha}{2\pi k_{B}}\mbox{,}
\]
i.e. the same result holds both for an accelerated observer at zero
temperature as for an inertial observer immersed in a thermal bath
at temperature $T$ . 

\begin{figure}[H]
\centering{}\includegraphics[scale=0.5]{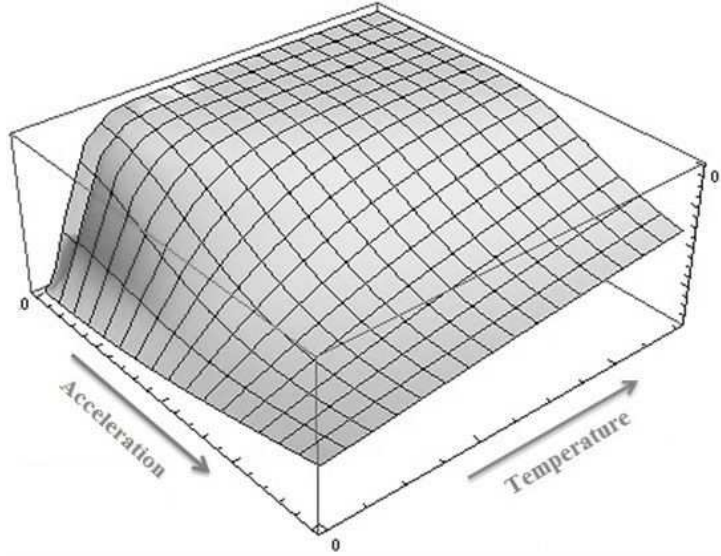}\protect\caption{The two-point function $g\left(\tau',\tau''\right)$}
\label{g.tau}
\end{figure}

Though, looking at the graph in Figure\ref{g.tau} to $g\left(\tau',\tau''\right)$
we found that the behaviour regarding the temperature and acceleration
are not the same. Note that, at zero temperature, as the acceleration
increases, $g\left(\tau',\tau''\right)$ grows almost linearly. However,
fixing a non-zero temperature, results that by increasing the acceleration
$g\left(\tau',\tau''\right)$ decreases. We conclude that the temperature
influences the way the proper acceleration behaves.

\section{Final Considerations\label{sec:Final-Considerations}}

In this paper, after introducing the original motivation behind the
so-called Unruh-Fulling-Davies-DeWitt effect through the response
function of a detector, which undergoes hyperbolic motion, we have
shown that the Thermofield Dynamics (FTD) formalism revealed itself
as a natural scenario for describing the behaviour of accelerated
systems at finite temperature. In such approach, the thermal states
belong to a Hilbert space in the number representation and describe
a microscopic system interacting with a \emph{reservoir}, leading
to equations of motion for both the two-level detector and the radiation
field, so that in the coarse-graining approach, the dynamics of these
systems consist of a Markovian process. Supposing that the sub-systems
reach a global thermal equilibrium at a common temperature, we have
shown that the local fluctuation due to admissible couplings, enlarging
the class of DeWitt detectors, lead to a master equation for the population
levels whose dynamics respect the fluctuation-dissipation theorem.
As a result, we arrived at the conclusion that the two-point correlation
function associated to the population levels depend on both the \emph{reservoir}
equilibrium temperature and the magnitude of the hyperbolic acceleration,
indicating the existence of a range of temperatures in which the population
of the excited state may also decrease at increasing acceleration
magnitudes \cite{Brenna:2015fga} and, therefore, Unruh and \emph{reservoir}
temperatures do not coincide, although the detailed balance condition
is respected. This is corroborated by the analysis of the vacuum fluctuation
and radiation reaction hermitian contributions to the energy variation
rates of the scalar field considered as a thermal \emph{reservoir}
by choosing a suitable operator ordering. So, we have found that the
evolution equation for an accelerated observer coincides with the
corresponding one for an inertial observer immersed in a thermal Unruh
bath if, and only if, one ascribes \emph{a priori }a zero value temperature
to the radiation field in Rindler coordinates. 

\appendix

\section{The Coarse-Graining approximation\label{chap:Condi=0000E7=0000E3o-de-Validade}}

In section \ref{sec:The-Master-Equation} the operator $\sigma_{A}\left(t"\right)$
has been replaced by $\sigma_{A}\left(t\right)$, which is equivalent
to expanding the master equation only to second order. If applied
recursively, this procedure will generate contributions to higher
orders, appearing terms of the type triple, quadruple commutators,
and so on. The formal convergence analysis of this series is of extreme
complexity, being out of the scope of this paper. However, we present
elements of plausibility, which indicate that, at least, the series
should asymptotically converge. The exact expression for the population
is 
\begin{equation}
\Delta\sigma_{A}\left(t\right)=\left(\frac{1}{i\hbar}\right)^{2}\int_{t}^{t+\Delta t}dt'\int_{t}^{t'}dt''Tr_{R}\left[V\left(t'\right),\left[V\left(t''\right),\rho\left(t''\right)\right]\right]\mbox{.}
\end{equation}

Integrating the equation for $\mbox{\ensuremath{\rho}}$ between $t$
and $t''$, 
\begin{equation}
\rho\left(t''\right)-\rho\left(t\right)=\frac{1}{i\hbar}\int_{t}^{t''}dt'''\left[V\left(t'''\right),\rho\left(t'''\right)\right]\mbox{,}
\end{equation}
and substituting in the above equation naturally gives rise to the
second and third order terms in the expansion, 
\begin{equation}
\begin{array}{c}
\Delta\sigma_{A}\left(t\right)=\left(\frac{1}{i\hbar}\right)^{2}\int_{t}^{t+\Delta t}dt'\int_{t}^{t'}dt''Tr_{R}\left[V\left(t'\right),\left[V\left(t''\right),\rho\left(t\right)\right]\right]\\
\\
+\left(\frac{1}{i\hbar}\right)^{3}\int_{t}^{t+\Delta t}dt'\int_{t}^{t'}dt''\int_{t}^{t''}dt'''Tr_{R}\left[V\left(t'\right),\left[V\left(t''\right),\left[V\left(t'''\right),\rho\left(t'''\right)\right]\right]\right]\mbox{.}
\end{array}
\end{equation}

Due to the shape of the potential, we find that the traces can be
factored into a part referring to the system $A$ and the other, $g\left(\tau\right)$,
to the system $R$ . This will allow us to ignore the higher order
terms. We must show that, after a time interval $\Delta t$ sufficiently
large, the three points correlation of the \emph{reservoir} observable
are more strongly suppressed. Projecting on the $R$ basis we have
\begin{equation}
g\left(t',t''\right)=Tr\left[\sigma_{R}R\left(t'-t''\right)R\right]=Z_{R}^{-1}\underset{m,n}{\sum}e^{-\beta E_{n}}\left\langle n\right|R^{S}\left|m\right\rangle \left\langle m\right|R^{S}\left|n\right\rangle e^{-i\omega_{nm}\left(t'-t''\right)}\mbox{,}
\end{equation}

\begin{equation}
\begin{array}{c}
g\left(t',t'',t'''\right)=Tr\left[\sigma_{R}R\left(t'-t''\right)RR\left(t'''-t''\right)\right]\\
\\
=Z_{R}^{-1}\underset{m,n,l}{\sum}e^{-\beta E_{n}}\left\langle n\right|R^{S}\left|m\right\rangle \left\langle m\right|R^{S}\left|l\right\rangle \left\langle l\right|R^{S}\left|n\right\rangle e^{-i\omega_{nm}\left(t'-t''\right)}e^{-i\omega_{nl}\left(t''-t'''\right)}\mbox{.}
\end{array}
\end{equation}

We cannot go further without additional information about the \emph{reservoir}.
However, the \emph{reservoir} has dense, almost continuous, spectrum
of energy so we assume that the different phases in $g\left(t',t''\right)$
cause a destructive interference as time increases. Thus, $g\left(t',t''\right)$
oscillates very rapidly and its contribution to the integral becomes
smaller. Considering two different time scales, such that 
\begin{equation}
T_{A}\gg\Delta t\gg\tau_{c}\mbox{,}
\end{equation}
where $\tau_{c}$ is the time correlation between observables of the
\emph{reservoir} and $T_{A}$ the time evolution associated with the
microscopic system, we have at second order, 
\begin{equation}
\left|\frac{\Delta\sigma_{A}}{\Delta t}\right|^{\left(2\right)}\sim\frac{1}{\Delta t}\frac{v^{2}}{\hbar^{2}}\sigma_{A}\left|\int_{t}^{t+\Delta t}dt'\int_{t}^{t'}dt''e^{-i\omega\left(t'-t''\right)}\right|\mbox{,}
\end{equation}
close to a given dominant frequency 
\begin{equation}
\omega_{mn}\equiv\omega\sim\tau_{c}^{-1}\mbox{.}
\end{equation}

In the expression, $v\equiv\left\langle V\left(\beta\right)\right\rangle $
is the typical value of the interaction potential at equilibrium,
the observable microscopic system is taken at a given time, around
which it does not vary appreciably within the integration intervals.
Performing the integral

\begin{equation}
\int_{t}^{t+\Delta t}dt'\int_{t}^{t'}dt''e^{-i\omega\left(t'-t''\right)}=\frac{2}{\omega}\int_{t}^{t+\Delta t}dt'e^{-i\frac{\omega}{2}\left(t'-t\right)}sin\left[\frac{\omega}{2}\left(t'-t\right)\right]
\end{equation}

\begin{equation}
\cong\frac{2}{\omega}e^{-i\omega\frac{\Delta t}{4}}sin\left[\frac{\omega\Delta t}{4}\right]\Delta t\mbox{;}
\end{equation}
therefore, 
\begin{equation}
\left|\frac{\Delta\sigma_{A}}{\Delta t}\right|^{\left(2\right)}\sim2\tau_{c}\frac{v^{2}}{\hbar^{2}}\sigma_{A}\sim\frac{\sigma_{A}}{T_{R}}\sim\sigma_{A}\left(\frac{\tau_{c}}{T_{R}}\right)\tau_{c}^{-1}\mbox{,}
\end{equation}
implying 
\begin{equation}
\frac{v^{2}}{\hbar^{2}}\tau_{c}^{2}\ll1\mbox{.}
\end{equation}

To third order we have 
\begin{equation}
\left|\frac{\Delta\sigma_{A}}{\Delta t}\right|^{\left(3\right)}\sim\frac{1}{\Delta t}\frac{v^{3}}{\hbar^{3}}\sigma_{A}\left|\int_{t}^{t+\Delta t}dt'\int_{t}^{t'}dt''\int_{t}^{t''}dt'''e^{-i\omega\left(t'-t'''\right)}\right|\mbox{;}
\end{equation}
resulting 
\begin{equation}
\left|\frac{\Delta\sigma_{A}}{\Delta t}\right|^{\left(3\right)}\sim4\frac{v^{3}}{\hbar^{3}}\tau_{c}^{2}\sigma_{A}\sim\left|\frac{\Delta\sigma_{A}}{\Delta t}\right|^{\left(2\right)}2\frac{v}{\hbar}\tau_{c}\ll\left|\frac{\Delta\sigma_{A}}{\Delta t}\right|^{\left(2\right)}\mbox{.}
\end{equation}
Then, in weakly coupled systems, the contribution of higher order
terms is less than the lower order ones, even though the series does
not converge uniformly, being only asymptotic.

\section{Derivative Coupling\label{sub:Acoplamento-derivativo}}

In this appendix we ask ourselves what could happen if instead of
\eqref{eq:hamiltoniano.intera=0000E7ao.audretsch}, we had a derivative
coupling. Thus, we propose a more general interaction given by

\begin{equation}
H_{I}=\mu R_{2}\left(\tau\right)\frac{1}{\omega_{0}^{n}}\frac{d^{n}}{d\tau^{n}}\phi\left(\tau\right)\mbox{,}\label{eq: nova.intera=0000E7ao}
\end{equation}
where the constant $\omega_{0}^{n}$ is introduced to ensure the correct
dimension. With the interaction \eqref{eq: nova.intera=0000E7ao},
we find new equations for the vacuum fluctuations and radiation reaction,
equivalent to the equations \eqref{eq:VFi.final} and \eqref{eq:RR.final},

\begin{equation}
\begin{array}{c}
\left\langle \frac{d}{d\tau}H_{A}\left(\tau\right)\right\rangle _{VF}=-\frac{\omega_{0}\mu^{2}}{\omega_{0}^{2n}}\intop_{\tau_{0}}^{\tau}d\tau'cos\left(\omega_{0}\left(\tau-\tau'\right)\right)\left\langle a\left|R_{3}^{f}\left(\tau_{0}\right)\right|a\right\rangle \times\\
\\
\left[i\left(2\lambda-1\right)\frac{d^{n}}{d\tau^{n}}\frac{d^{n}}{d\tau'^{n}}\chi^{F}\left(\tau,\tau'\right)+\frac{d^{n}}{d\tau^{n}}\frac{d^{n}}{d\tau'^{n}}C^{F}\left(\tau,\tau'\right)\right]\mbox{,}
\end{array}\label{eq:VF.derivativo}
\end{equation}

$ $

\begin{equation}
\begin{array}{c}
\left\langle \frac{d}{d\tau}H_{A}\left(\tau\right)\right\rangle _{RR}=\frac{\omega_{0}\mu^{2}}{\omega_{0}^{2n}}\intop_{\tau_{0}}^{\tau}d\tau'\left[\frac{d^{n}}{d\tau^{n}}\frac{d^{n}}{d\tau'^{n}}\chi^{F}\left(\tau,\tau'\right)\right]\times\\
\\
\left[\frac{1}{2}\,sin\left(\omega_{0}\left(\tau-\tau'\right)\right)+i\left(2\lambda-1\right)cos\left(\omega_{0}\left(\tau-\tau'\right)\right)\left\langle a\left|R_{3}^{f}\left(\tau_{0}\right)\right|a\right\rangle \right]\mbox{.}
\end{array}\label{eq:RR.Derivativo}
\end{equation}

Choosing the symmetrical order, $\lambda=\frac{1}{2}$, we eliminate
the divergences both in \eqref{eq:VF.derivativo} and \eqref{eq:RR.Derivativo}
. Then, by solving the integrals present in \eqref{eq:VF.derivativo}
and \eqref{eq:RR.Derivativo}, we recover \eqref{eq:VF.rindleri.final.simetrico}
and \eqref{eq:RR.rindler.final.simetrico} for the non-derivative
case. We thus observe that both results are independent of the coupling
order, always giving the same energy variations.

\section{Field Energy Variation \label{chap:Varia=0000E7=0000E3o-de-Energia}}

In section \ref{sub:Master-2-level}, we derived an equation for the
energy variation rate of the atomic system due to vacuum fluctuations
and radiation reaction. In this appendix, we show that the energy
balance is respected, observing that the energy variation of the field
is equal, in module, to that of the atomic system. We start from the
Heisenberg equation \eqref{eq:eq.heisenberg}
\begin{equation}
\frac{d}{d\tau}H_{F}\left(\tau\right)=i\left[H_{I}\left(\tau\right),H_{F}\left(\tau\right)\right]=i\mu R_{2}\left(\tau\right)\left[\phi\left(\tau\right),H_{F}\left(\tau\right)\right]\mbox{.}
\end{equation}

As previously said, it is advantageous to symmetrizing the operators
in the evolution equation,

\begin{equation}
\begin{array}{c}
\frac{d}{d\tau}H_{F}\left(\tau\right)=\mu\frac{1}{2}\left\{ \left(\partial_{\tau}\phi^{f}\left(\tau\right)\right)\left(R_{2}^{f}\left(\tau\right)+R_{2}^{S}\left(\tau\right)\right)+\left(R_{2}^{f}\left(\tau\right)+R_{2}^{S}\left(\tau\right)\right)\partial_{\tau}\phi^{f}\left(\tau\right)\right\} \\
\\
+\mu\frac{1}{2}\left\{ \left(\partial_{\tau}\phi^{S}\left(\tau\right)\right)R_{2}^{f}\left(\tau\right)+R_{2}^{f}\left(\tau\right)\partial_{\tau}\phi^{S}\left(\tau\right)\right\} .
\end{array}
\end{equation}

We separate again the free of the field responsible for the vacuum
fluctuation,

\begin{equation}
\left(\frac{d}{d\tau}H_{F}\left(\tau\right)\right)_{VF}=\mu\frac{1}{2}\left\{ \left(\partial_{\tau}\phi^{f}\left(\tau\right)\right)\left(R_{2}^{f}\left(\tau\right)+R_{2}^{S}\left(\tau\right)\right)+\left(R_{2}^{f}\left(\tau\right)+R_{2}^{S}\left(\tau\right)\right)\partial_{\tau}\phi^{f}\left(\tau\right)\right\} ,\label{eq:VF.campo}
\end{equation}
as well as the interaction part, responsible for the radiation reaction,

\begin{equation}
\left(\frac{d}{d\tau}H_{F}\left(\tau\right)\right)_{RR}=\mu\frac{1}{2}\left\{ \left(\partial_{\tau}\phi^{S}\left(\tau\right)\right)R_{2}^{f}\left(\tau\right)+R_{2}^{f}\left(\tau\right)\partial_{\tau}\phi^{S}\left(\tau\right)\right\} \mbox{.}\label{eq:RR.campo}
\end{equation}

Taking the expectation value in the field vacuum $\left|0\right\rangle $
and the atomic state $\left|a\right\rangle $, we obtain 
\[
\left\langle \frac{d}{d\tau}H_{F}\left(\tau\right)\right\rangle _{VF}=-i\mu^{2}\frac{1}{2}\intop_{\tau_{0}}^{\tau}d\tau'\left\langle a\left|\left[R_{2}^{f}\left(\tau\right),R_{2}^{f}\left(\tau'\right)\right]\right|a\right\rangle \partial_{\tau}\left\langle 0\left|\left[\phi^{f}\left(\tau\right),\phi^{f}\left(\tau'\right)\right]_{+}\right|0\right\rangle \mbox{,}
\]

\[
\left\langle \frac{d}{d\tau}H_{F}\left(\tau\right)\right\rangle _{RR}=-i\mu^{2}\frac{1}{2}\intop_{\tau_{0}}^{\tau}d\tau'\left\langle a\left|\left[R_{2}^{f}\left(\tau\right),R_{2}^{f}\left(\tau'\right)\right]_{+}\right|a\right\rangle \partial_{\tau}\left\langle 0\left|\left[\phi^{f}\left(\tau\right),\phi^{f}\left(\tau'\right)\right]\right|0\right\rangle \mbox{,}
\]
where we can recognize the correlation functions and susceptibility
\eqref{eq:func.correl.campo}, \eqref{eq:func.suscep.campo}, \eqref{eq:func.correl.atomo}
and \eqref{eq:func.susep.atomo}, 

\begin{equation}
\left\langle \frac{d}{d\tau}H_{F}\left(\tau\right)\right\rangle _{VF}=2\mu^{2}\intop_{\tau_{0}}^{\tau}d\tau'\chi^{A}\left(\tau,\tau'\right)\partial_{\tau}C^{F}\left(\tau,\tau'\right)\mbox{,}
\end{equation}

\begin{equation}
\left\langle \frac{d}{d\tau}H_{F}\left(\tau\right)\right\rangle _{RR}=2\mu^{2}\intop_{\tau_{0}}^{\tau}d\tau'C^{A}\left(\tau,\tau'\right)\partial_{\tau}\chi^{F}\left(\tau,\tau'\right)\mbox{.}
\end{equation}

Finally, replacing the functions found for $C^{A}$, $C^{F}$, $\chi^{A}$
and $\chi^{F}$ , for the asymptotic case,$\tau_{0}\rightarrow-\infty$,
results

\begin{equation}
\left\langle \frac{d}{d\tau}H_{F}\left(\tau\right)\right\rangle _{VF}=\frac{\mu^{2}}{4\pi^{2}}\underset{k=-\infty}{\overset{\infty}{\sum}}\intop_{0}^{\infty}du\,sin\left(\omega_{0}u\right)\left[\frac{1}{\left(u-i\epsilon2+i\frac{2\pi}{\alpha}k\right)^{3}}+\frac{1}{\left(u+i\epsilon2+i\frac{2\pi}{\alpha}k\right)^{3}}\right]\left\langle a\left|R_{3}\right|a\right\rangle \mbox{,}
\end{equation}

\begin{equation}
\left\langle \frac{d}{d\tau}H_{F}\left(\tau\right)\right\rangle _{RR}=\frac{\mu^{2}}{8\pi^{2}i}\underset{k=-\infty}{\overset{\infty}{\sum}}\intop_{0}^{\infty}du\,cos\left(\omega_{0}u\right)\left[\frac{1}{\left(u-i\epsilon2+i\frac{2\pi}{\alpha}k\right)^{3}}-\frac{1}{\left(u+i\epsilon2+i\frac{2\pi}{\alpha}k\right)^{3}}\right]\mbox{;}
\end{equation}
solving the integrals, one finds the same absolute values, obtained
in \eqref{eq:VF.rindleri.final.simetrico} and \eqref{eq:RR.rindler.final.simetrico},
ensuring that the energy balance is respected.

\newpage

\end{document}